\documentclass{article}
\usepackage{algorithm}
\usepackage{algpseudocode}
\usepackage[utf8]{inputenc}
\usepackage[T1]{fontenc}   
\usepackage{multirow}
\usepackage{tikz}
\usepackage{PRIMEarxiv}
\usepackage[sorting=none]{biblatex}
\bibliography{references}
\usepackage[inkscapelatex=false]{svg}
\usepackage[frozencache, cachedir=.]{minted}
\usepackage{hyperref}       
\usepackage{url}            
\usepackage{booktabs}       
\usepackage{amsfonts}       
\usepackage{nicefrac}       
\usepackage{microtype}      
\usepackage{lipsum}
\usepackage{fancyhdr}       
\usepackage{graphicx}       
\usepackage{amsmath} 
\usepackage{multicol}
\usepackage{caption}
\graphicspath{{media/}}     
\usepackage{todonotes}
\title{{\sf PC-Gym}: Benchmark Environments for Process Control Problems}

\author{
  Maximilian Bloor$^{1}$, José Torraca$^{1,3}$,  Ilya Orson Sandoval$^{1}$, Akhil Ahmed$^{1}$, Martha White$^{4}$\\ \textbf{Mehmet Mercangöz$^{1}$, Calvin Tsay$^{2}$, Ehecatl Antonio Del Rio Chanona$^{1}$, Max Mowbray$^{1}$}\footnote{Corresponding Author} \\
   $^{1}$Department of Chemical Engineering, Imperial College London, London, SW7 2AZ, United Kingdom \\
   $^{2}$Department of Computing, Imperial College London, London, SW7 2AZ, United Kingdom \\
  $^{3}$School of Chemistry, Universidade Federal do Rio de Janeiro, Rio de Janeiro, 21941-909, Brazil\\
  $^{4}$Alberta Machine Intelligence Institute (Amii), Department of Computing Science, University of Alberta\\
  \texttt{\{max.bloor22, m.mowbray\}@imperial.ac.uk} \\
}

\begin{document}
\maketitle

\begin{abstract}
{\sf PC-Gym} is an open-source tool for developing and evaluating reinforcement learning (RL) algorithms in chemical process control. It features environments that simulate various chemical processes, incorporating nonlinear dynamics, disturbances, and constraints. The tool includes customizable constraint handling, disturbance generation, reward function design, and enables comparison of RL algorithms against Nonlinear Model Predictive Control (NMPC) across different scenarios. Case studies demonstrate the framework's effectiveness in evaluating RL approaches for systems like continuously stirred tank reactors, multistage extraction processes, and crystallization reactors. The results reveal performance gaps between RL algorithms and NMPC oracles, highlighting areas for improvement and enabling benchmarking.
By providing a standardized platform, PC-Gym aims to accelerate research at the intersection of machine learning, control, and process systems engineering. By connecting theoretical RL advances with practical industrial process control applications, offering researchers a tool for exploring data-driven control solutions.
\end{abstract}

\keywords{Reinforcement Learning \and Process Control}

\section{Introduction}
\label{sec:intro}
In the field of control systems, many problems are addressed using discrete-time approaches, where the system state is sampled (or estimated from output measurements) and control actions are applied at predefined intervals. This paradigm has been successfully employed in various domains, such as robotics, buildings, and automotive systems, enabling the development of effective control strategies. In recent years, model-free reinforcement learning (RL) has become a powerful data-driven framework for such problems. RL allows agents to learn optimal control policies through interaction with their environment, adapting to complex, unknown dynamics and uncertainties~\cite{suttonRL}. 

Chemical processes exhibit dynamics and constraints that set them apart from traditional RL settings. These processes often involve multiple interacting variables, nonlinear behavior, and time-varying parameters, making them particularly difficult to control using standard RL techniques \cite{nian_review_2020, shin_reinforcement_2019}. Furthermore, chemical processes are often subject to strict safety, environmental, and economic constraints, which must be carefully considered in the control strategy. Violating these constraints can lead to severe consequences, such as equipment damage, product quality degradation, or even human harm.

Due to these complexities, several works have explored the development and application of RL algorithms for chemical process control. Early applications of RL in chemical process control focused on model-free approaches for tracking control and optimization in fed-batch bioreactors \cite{kaisare2003simulation, peroni2005optimal, wilson1997neuro} which present unique challenges due to their nonstationary nature and significant run-to-run variabilities \cite{YOO2021108}. For batch processes, several works have explored using RL, for phase segmentation strategies where separate RL agents handle different operational phases of the process \cite{yoo_reinforcement_2021}. Some approaches have focused on integrating imitation learning with RL to learn from process demonstrations and expert knowledge \cite{faria_where_2022}. Efforts have also been made to improve RL training stability through techniques like Monte Carlo learning and reward function design that considers process constraints \cite{yoo_reinforcement_2021, bao_deep_2021}. \textcite{PETSAGKOURAKIS202235} introduced a chance-constrained policy optimization algorithm that guarantees the satisfaction of joint chance constraints with high probability. Efforts to improve the scalability of RL algorithms by reducing control-space dimensionality have also shown potential in plantwide control problems~\cite{zhu2020scalable}. Works have explored incorporating control structures common in chemical processes such as proportional-integral-derivative (PID) controllers. \textcite{lawrence2022deep} directly parameterized the RL policy as a PID controller instead of using a deep neural network. This allows the RL agent to improve the controller's performance while leveraging existing PID control hardware, demonstrating that industry can utilize actor-critic RL algorithms without additional hardware or sacrificing the interpretability often lost when using deep neural networks. \textcite{CIRL} integrated the neural network and PID controller into a singular RL agent which improved the sample efficiency of the agent compared to a standard deep RL agent due to the embedded control structure. These works all rely on creating discrete control environments, which has motivated the attention of other scientific communities to create open-source accessible environments.

\subsection{Related Works}
The development of reinforcement learning (RL) algorithms relies heavily on the availability of easily implementable environments that allow for efficient experimentation and iteration. These environments are crucial for benchmarking algorithms along with proper empirical practice as shown in work by \textcite{patterson2024empiricaldesignreinforcementlearning} and \textcite{jordan2024positionbenchmarkinglimitedreinforcement}. Frameworks such as Gymnasium (formerly known as Gym) by Farama-Foundation \cite{towers_gymnasium_2023} and OpenAI's Gym \cite{gym} have been instrumental in providing standardized discrete-time environments for RL research, offering a wide range of control problems, games, and simulations. However, these frameworks lack the latest parallelization methods for efficiently learning effective control policies and developing algorithms. To overcome this limitation, several libraries, such as Gymnax \cite{gymnax2022github}, Brax \cite{brax2021github}, Jumanji \cite{bonnet2024jumanji}, and PGX \cite{koyamada2023pgx}, have been developed to provide vectorized RL environments in JAX~\cite{jax2018github}, enabling parallel simulations on Graphics Processing Units while maintaining the discrete-time nature of the environments. While these frameworks cover a variety of control problem types, there is still a need for domain-specific environments. Consequently, various natural sciences and engineering communities have started creating custom environments in the same format to simulate their specific problems, such as OR-Gym \cite{HubbsOR-Gym} for operational research and ChemGymRL \cite{chemgymrl} for chemical discovery. Despite these advancements, there remains a gap in the availability of process control environments with specific functionality tailored to creating discrete-time process control problems.

\subsection{Contributions}
This work presents {\sf PC-Gym}, a comprehensive benchmarking suite and development tool for process control algorithms in the chemical industry. The key contributions of this work are as follows:
\begin{enumerate}
    \item \textbf{Chemical process control problems}: {\sf PC-Gym} encompasses a variety of unit operations, each considered to represent realistic process dynamics and objectives.
    \item \textbf{Disturbance generation}: The suite includes functionality for creating and incorporating disturbances, allowing researchers to test the robustness of their algorithms under various perturbations.
    \item \textbf{Constraint handling}: {\sf PC-Gym} provides mechanisms for incorporating process constraints, encouraging the development of algorithms that respect real-world limitations.
    \item \textbf{Policy visualization and evaluation tools}: The suite offers tools for visualizing and evaluating the performance of control policies against a model-based oracle, facilitating the analysis and interpretation of algorithm behavior.
\end{enumerate}
By providing a set of chemical process control scenarios, disturbance, and constraint generation capabilities, and policy visualization and evaluation tools, the suite serves as a valuable resource for researchers and practitioners. The {\sf PC-Gym} library including code for reproducing the case studies within this work is available at \href{https://github.com/MaximilianB2/pc-gym}{{\fontfamily{qcr}\selectfont
https://github.com/MaximilianB2/pc-gym}}.

The remainder of this paper is organized as follows: Section \ref{sec:background} gives a brief background on reinforcement learning and a formal representation of how {\sf PC-Gym} represents dynamical systems, Section \ref{sec:software_implementation} describes the software implementation of {\sf PC-Gym},  Section \ref{sec:controlproblems} describes the dynamical systems modeled in {\sf PC-Gym} and benchmarks the performance of common RL algorithms. Finally, Sections \ref{sec:disturbance}, \ref{sec:constraints}, and \ref{sec:rewardfunc} demonstrate {\sf PC-Gym}'s disturbance generation, constraint, and reward function customization functionality respectively. 

\section{Background}
\label{sec:background}
\subsection{Dynamical systems}\label{sec:dynamical_systems}
In this work, we assume discrete-time stochastic descriptions of dynamical systems. This is primarily due to the discrete-time nature of modern instrumentation systems, which makes such a framework practical. Specifically, the state transition is assumed as,
\begin{equation}\label{eq:dynamical_system}
    X_{t+1} \sim p(\textbf{x}_{t+1} | \textbf{x}_t, \textbf{u}_t)\,,
\end{equation}
where $\textbf{x}_t \in \mathcal{X} \subseteq \mathbb{R}^{n_x}$ indicates the system states; $\textbf{u}_t \in \mathcal{U}\subseteq\mathbb{R}^{n_u}$ represents the system's control inputs; the subscript indicates the quantity is measured at timestep $t$; and $p(\cdot)$ represents a conditional probability distribution. We hypothesize that the conditional probability density could be a representation of a nonlinear state space model,
\begin{equation*}
    \textbf{x}_{t+1} = f(\textbf{x}_t, \textbf{u}_t, \textbf{w}_t)\,,
\end{equation*}
with stochasticity as the result of general uncertain parameters, $\textbf{w}_t \in \mathcal{W}\subseteq\mathbb{R}^{n_w}$, indicating disturbance or model parameters, with $\mathcal{W} = \text{supp}(p(\textbf{w}))$, being the support of its distribution.   This work aims to characterize a set of transition density functions $p(\cdot) \in \mathcal{P}$ to represent meaningful process control problems. Each transition function is parameterized in general state-space form, with means to specify different disturbances as the user deems appropriate.
\subsection{Learning-based Control}\label{sec:learning-based-control}
Given a dynamical system as defined in Equation \ref{eq:dynamical_system}, an initial state is sampled from a distribution $X_0 \sim p_0(\textbf{x}_0)$ where $p_0(\cdot)$ is an assumed initial state distribution. The control problem is defined by the control inputs $\textbf{u}_t$, which maximize a sequence of rewards, allocated by a function, $r_{t+1}: \mathcal{X} \times \mathcal{U} \rightarrow \mathbb{R}$. Learning-based control aims to identify a state feedback policy based on data collected from a dynamical system over a fixed horizon $T$. Here, the control problem is approximated with a fixed horizon to create an episodic learning approximation to the infinite horizon task at hand. Therefore, the objective is to find the optimal control policy $\pi^*: \mathcal{X} \rightarrow \mathcal{U}$ that maps state to control actions, maximizing the expected cumulative reward.
\begin{align*}
    \max_{\pi} \hspace{2mm}&\mathbb{E}_{\pi}\left[\sum^{T-1}_{t=0}\gamma^t r_{t+1}(X_t,U_t)\right] \\
s.t.\hspace{0.5cm} 
& X_0 \sim p_0(\textbf{x}_0) \\
&  X_{t+1} \sim p(\textbf{x}_{t+1} | \textbf{x}_t, \textbf{u}_t)\\
& U_t \sim \pi(\textbf{u}_t \mid \textbf{x}_t)  \\
&t \in \{0, \ldots, T-1\}\,,
\end{align*}
with $\mathbb{E}_\pi[\cdot]$ denoting the expectation under the policy $\pi$. 
\subsection{Reinforcement Learning}

\begin{figure}[ht!]
    \centering
    \includegraphics[clip, trim=1.1in 3.1in 2.9in 0.5in, width=0.5\textwidth]{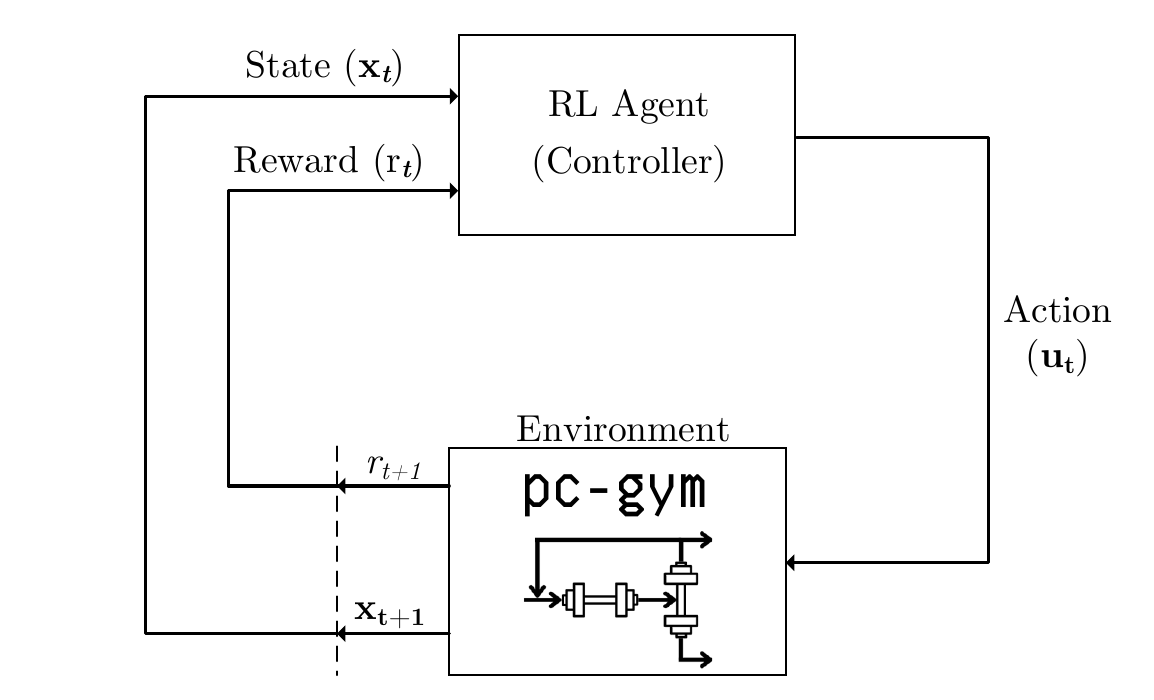}
    \caption{The reinforcement learning framework}
    \label{fig:rl-framework}
\end{figure}

A popular method to solve the control problem described in the previous section is reinforcement learning (RL), where the major assumption applied is that the system is Markovian and that the problem is modeled as a Markov Decision Process (MDP). Figure \ref{fig:rl-framework} describes the MDP formulation as the interaction between a control policy (agent) and a discrete-time stochastic decision process (environment). The control policy observes the state of the process and inputs a control decision to the environment, the environment then transitions to a new state and returns a reward based on a reward function.

Many types of algorithms have emerged to find an approximate optimal policy in the RL framework. In the context of chemical process control, three advanced algorithms have shown particular promise: Proximal Policy Optimization (PPO), Soft Actor-Critic (SAC), and Deep Deterministic Policy Gradient (DDPG). These algorithms are well-suited to handle continuous action spaces and complex dynamics often encountered in process control problems. Central to many RL algorithms is the Q-function, also known as the state-action value function. Mathematically, it can be expressed as the expectation of a discounted sum of scalar random variables:

\begin{equation}
   Q^\pi(\textbf{x}_t, \textbf{u}_t) = \mathbb{E}_\pi\left[\sum_{k=t}^{T-1} \gamma^{k-t} r_{k+1}(X_k, U_k)
    \middle| X_{t} = \textbf{x}_t, U_{t} = \textbf{u}_t\right]\,.
\end{equation}

This formulation explicitly shows the Q-function as the expected sum of discounted future rewards, starting from state $\textbf{x}_t$ and taking control $\textbf{u}_t$, then following the policy for all subsequent steps. Another key concept is the value function $V^\pi(\textbf{x}_t)$, which represents the expected return starting from state $\textbf{x}_t$ and following the policy $\pi$. It can be defined as,

\begin{equation}
    V^\pi(\textbf{x}_t) = \mathbb{E}_\pi\left[\sum_{k=t}^{T-1} \gamma^{k-t} r_{k+1}({X}_{k}, U_{k}) \middle| X_{t} = \textbf{x}_t\right],
\end{equation}

where $\gamma\in[0,1]$ is the discount factor and $U_k$ represents the control action generated by policy $\pi$ at each timestep, $k\in \{t, \ldots, T-1\}$. The value function helps in evaluating how ``good'' it is to be in a particular state, considering the long-term rewards that can be obtained from that state under the current policy. The advantage function, which is used in many RL algorithms, combines the Q-function and the value function:

\begin{equation}\label{eq:advantaje}
   A^\pi(\textbf{x}_t, \textbf{u}_t) = Q^\pi(\textbf{x}_t, \textbf{u}_t) - V^\pi(\textbf{x}_t)\,.
\end{equation}

This function estimates the difference in performance between taking a given control $\textbf{u}_t$, in state $\textbf{x}_t$, and following the policy thereafter, rather than simply following the policy. This provides an effective quantity for approximate policy improvement steps.

In practice, these functions are typically parameterized using neural networks due to their ability to handle continuous state and control spaces. The Q-function can be parameterized by $\theta$ as $Q_\theta(\textbf{x}_t, \textbf{u}_t)$, and the policy as $\pi_\phi(\textbf{u}_t|\textbf{x}_t)$ with parameters $\phi$. The parameterized policy can be either deterministic, directly outputting controls $\textbf{u}_t = \pi_\phi(\textbf{x}_t)$, or stochastic, outputting a probability density over controls. For continuous control problems in chemical processes, the policy network often outputs the mean and standard deviation of a Gaussian distribution:
\begin{equation}
    \pi_\phi(\textbf{u}_t|\textbf{x}_t) = \mathcal{N}(\boldsymbol{\mu}_\phi(\textbf{x}_t), diag(\boldsymbol{\sigma}_\phi(\textbf{x}_t))),
\end{equation}
where $\boldsymbol{\mu}_\phi:\mathbb{R}^{n_x}\rightarrow \mathbb{R}^{n_u}$ and $\boldsymbol{\sigma}_\phi:\mathbb{R}^{n_x}\rightarrow \mathbb{R}^{n_u}$ are neural network outputs\footnote{It is worth noting in deployment, Gaussian policies may be transformed into deterministic controllers by selecting the control corresponding to the mean. Control bounds may be enforced by a hyperbolic tangent function \cite{haarnoja2018soft} or simply clipping the policy output.}. The value function is similarly parameterized as $V_\psi(\textbf{x}_t)$ with parameters $\psi$. These parameterizations transform the reinforcement learning problem into one of finding optimal parameters $\phi$, $\theta$ and $\psi$. These are generally estimated through gradient-based optimization with gradients estimated through the policy gradient theorem \cite{sutton_policy_nodate} in the case of $\phi$, and variations of temporal difference (TD) learning in the case of $\theta$ \cite{watkins1992q} and $\psi$ \cite{boyan1999least}. The policy gradient provides a direction in the parameter space to maximize the expected returns for the policy. TD learning provides a means to identify a bootstrapped estimate of the squared error associated with a Q-function or value function approximation under a given behavior \cite{suttonRL} or target policy \cite{watkins1992q}. Taking the negative of the gradient with respect to this estimate provides a direction in the parameter space to improve the approximation. In the following, we give a brief description of three learning algorithms that have proved effective policy optimizers within the literature.

\subsubsection{Soft Actor-Critic (SAC)}
SAC is an off-policy actor-critic method that incorporates entropy regularization into the objective function~\cite{haarnoja2018soft}. Actor-critic methods parameterize both a policy (actor) and a Q-function or a value function (critic). The SAC algorithm employs both a value function and a Q-function together with a policy. Additionally, it uses experience replay and target networks~\cite{mnih_human-level_2015} for both the Q-function and value function approximation, which are updated with soft updates. It aims to learn a policy that maximizes both the expected return and the entropy of the policy. In an episodic setting, the objective function $J(\pi)$ represents the expected return of the policy and is defined as:
\begin{equation}\label{eq:real_obj}
J(\pi_\phi) = \mathbb{E}_{\pi_\phi}\left[\sum_{t=0}^{T-1} \gamma^t r_{t+1}(X_t, U_t)\right]\,.
\end{equation}
SAC augments this objective with an entropy term:
\begin{equation}
J_{SAC}(\pi_\phi) = J(\pi_\phi) + \alpha H(\pi_\phi)\,,
\end{equation}
where $H(\pi_\phi)$ is the expected entropy of the policy given the associated state distribution, and $\alpha$ is a temperature parameter that can be automatically adjusted through gradient descent to maintain a target entropy level. SAC learns a stochastic policy that parameterizes the moments of a Gaussian distribution conditional to the current state, which can be beneficial in chemical process environments with multiple optimal solutions or where exploration is crucial \cite{pmlr-v97-ahmed19a}.

\subsubsection{Deep Deterministic Policy Gradient (DDPG)}
DDPG is an off-policy actor-critic algorithm designed specifically for continuous control spaces, making it particularly suitable for process control tasks~\cite{DDPG}. For complete implementation details, readers are referred to the original paper. The algorithm implements experience replay and employs target networks for both policy and Q-function with soft updates. The networks are fully connected, where the policy outputs deterministic controls. The actor is updated using the deterministic policy gradient:
\begin{equation}
\nabla_{\phi} J = \mathbb{E}_{\textbf{x}_t \sim \rho^\beta}\left[\nabla_\textbf{u} Q(\textbf{x},\textbf{u};\theta)\mid_{{\textbf{x}=\textbf{x}_t,\textbf{u}=\mu(\textbf{x}_t;\phi)}} \nabla_{\phi} \boldsymbol{\mu}(\textbf{x}; \phi)\mid_{{\textbf{x}=\textbf{x}_t}}\right]
\end{equation}
where $\boldsymbol{\mu}(\textbf{x};\phi)$ is the deterministic policy and $Q(\textbf{x},\textbf{u};\theta)$ is the Q-function. The term $\rho^\beta$ denotes the state distribution under the behavior policy $\beta$. This behavior policy adds Ornstein-Uhlenbeck noise to the deterministic controls for exploration. The state distribution $\rho^\beta$ represents the probability distribution of states encountered when following this behavior policy.

\subsubsection{Proximal Policy Optimization (PPO)}
PPO builds on Trust Region Policy Optimization (TRPO)~\cite{schulman_trust_2015}, presenting a more computationally efficient approach to stable policy updates~\cite{schulman2017proximalpolicyoptimizationalgorithms}. While TRPO enforces a trust region through a KL-divergence constraint, PPO achieves similar stability through a simpler clipped objective function. Unlike SAC and DDPG, PPO does not use experience replay or target networks, instead collecting trajectories using the current policy for updates. It uses a policy-network that outputs the mean and standard deviation of a Gaussian policy and a value function network and introduces a clipped surrogate objective function:
\begin{equation}
    L^{CLIP}(\phi_j, \phi_{j-1}) = \mathbb{E}_{(\textbf{x},\textbf{u}) \sim \pi_{\phi_{j-1}}} \left[ \min\left(\frac{\pi_{\phi_j}}{\pi_{\phi_{j-1}}}A_{\psi}(\textbf{x}, \textbf{u}), \text{clip}\left(\frac{\pi_{\phi_j}}{\pi_{\phi_{j-1}}}, 1-\epsilon, 1+\epsilon\right)A_{\psi}(\textbf{x}, \textbf{u})\right)\right]\,,
\end{equation}
where $\frac{\pi_{\phi_j}}{\pi_{\phi_{j-1}}}$ is the probability ratio between the new and old policies, $A_{\psi}(\textbf{x}, \textbf{u})$ is the estimated advantage function calculated using Generalized Advantage Estimation (GAE)\footnote{Although the definition provided by Eq. \ref{eq:advantaje} indicates that the advantage estimate would require parameterization of both the state-action value function and the value function, one only needs the latter. Please see the cited work \cite{schulman_high-dimensional_2018} for more information.}~\cite{schulman_high-dimensional_2018}, $j$ is the policy update step and $\epsilon$ is a hyperparameter that controls the clipping range. This objective function is a surrogate to Equation \ref{eq:real_obj}, which imposes an approximate trust-region and ensures that the policy updates are neither too large nor too small, leading to more stable learning. This is particularly beneficial for highly nonlinear chemical processes.

Throughout this paper, we will use PPO, SAC, and DDPG to demonstrate the capabilities of {\sf PC-Gym}. These algorithms have been chosen for their effectiveness in dealing with continuous control spaces and their ability to handle the complexities often encountered in chemical process control problems. By using {\sf PC-Gym}, we aim to provide insights into the relative strengths and weaknesses of these algorithms across different process control scenarios.
\subsection{Constrained Reinforcement Learning}

In many real-world applications, particularly in chemical process control, the learning-based control problem must consider operational constraints. These constraints may represent safety limits, equipment capabilities, or product quality requirements. To incorporate these constraints, we can reformulate the learning-based control problem introduced in Section \ref{sec:learning-based-control} into an example constrained reinforcement learning formulation:

\begin{align*}
    \max_{\pi} \hspace{2mm}&\mathbb{E}_{\pi}\left[\sum^{T-1}_{t=0} \gamma^tr_{t+1}(X_t,U_t)\right] \\
s.t.\hspace{0.5cm}&X_0 \sim p_0(\textbf{x}) \\
&  X_{t+1} \sim p(\textbf{x}_{t+1} | \textbf{x}_t, \textbf{u}_t)\\ 
& \mathbb{P}\left[{g}_i(X_t) \leq 0\right] \geq 1-\alpha_{i,t}, \quad i \in \{1, \ldots, n_g\}\\
& t \in \{0, \ldots, T\}\,,
\end{align*}

where $g_i(\cdot)$ represents the $i$-th inequality constraint. The total number of constraints imposed over the time horizon is $n_g \times T$, therefore the problem imposes an individual chance constraint for each constraint function at each time index, given the uncertainty of the state over the time horizon. Explicitly, this allows for constraint violation with a given probability, $\alpha_{i,t}$. It is worth noting that one may also allow a probability for violation of the constraints jointly. This is known as a joint chance constraint \cite{paulson_smpc, PETSAGKOURAKIS202235}. In practice, this is generally enabled by tuning the individual constraints imposed, for example through constraint tightening \cite{PAN2021107462}. Recent work has shown that constraint-tightening approaches can effectively guarantee satisfaction of joint chance constraints with high probability while maintaining good control performance.

Various approaches can be employed to solve constrained RL problems. These include: Lagrangian methods \cite{BORKAR2005207, peng_separated_2021}, which transform the constrained optimization problem into an unconstrained one by incorporating constraints into the objective function through Lagrange multipliers; reward shaping techniques that modify the reward function to incorporate penalties for constraint violations \cite{peng_separated_2021}; and safe policy optimization algorithms like Constrained Policy Optimization (CPO) that directly optimize policies while ensuring constraint satisfaction throughout the learning process \cite{pmlr-v70-achiam17a}. The Separated Proportional-Integral Lagrangian (SPIL) approach combines the benefits of both penalty and Lagrangian methods to achieve better performance while maintaining safety guarantees \cite{peng_separated_2021}.

A key metric in evaluating constrained RL algorithms is the joint probability of constraint violation. Here, the constraints are treated jointly to provide a single metric to evaluate constraint satisfaction. Formally, for a given policy $\pi$ the joint probability of constraint violation can be defined as:
\begin{equation}
    P_{\text{violation}}(\pi) = 1- \mathbb{P}\left( \bigcap_{t=0}^T \{\textbf{x}_t \in \mathbb{X}_t \}\right)\,,
\end{equation}
with, 
\begin{equation*}
    \mathbb{X}_t = \{ \textbf{x}_t \in \mathbb{R}^{n_x} \mid g_i(\textbf{x}_t) \leq 0, \ \forall i \in \{1, \ldots, n_g\}\}\,.
\end{equation*}

In practice, this probability is often estimated empirically according to an empirical cumulative distribution function defined as the result of multiple episodes or rollouts~\cite{petsagkourakis2022chance}.

In the context of chemical process control, constrained RL can be particularly valuable for ensuring that learned control policies maintain safe operating conditions, adhere to equipment limitations, and satisfy product quality constraints. The ability to incorporate such constraints directly into the learning process, while quantifying and minimizing the risk of violations, can significantly enhance the practical applicability of RL in industrial settings, hence user-defined constraints and constraint violation information are available for the user to incorporate into their algorithm. Within this work, only reward shaping with constraint violation information is shown since our aim is to demonstrate the flexibility of the package.
\subsection{Policy Performance and Reproducibility Measures}
\label{sec:policy_eval_background}
In reinforcement learning (RL), assessing the performance of trained policies requires metrics that capture both the performance and the reproducibility of the learned policy. Dispersion metrics play a crucial role in quantifying the variability of policy performance across multiple training runs. While the expected return provides insight into a policy's average performance, metrics like Median Absolute Deviation (MAD) and standard deviation offer valuable information about the dispersion of returns from a policy. These dispersion metrics are particularly important in domains with inherent stochasticity, such as chemical process control, where reproducible behavior is essential.

Given the inherent stochastic nature of chemical systems and RL policies, the same control policy $\pi$ can perform differently across runs, which we define as reproducibility \cite{flageat2024expectedreturnaccountingpolicy}. Evaluating both the expected return $J(\pi)$ and reproducibility of policies is crucial. {\sf PC-Gym} addresses this by reporting a performance measure, such as the median return $\tilde{J}(\pi)$, along with a chosen dispersion metric across multiple repetitions in a fixed chemical process control environment. This dispersion metric can either be the median absolute deviation $\text{MAD}(\pi)$ or the standard deviation $J_\sigma(\pi)$ of returns. For instance, $\text{MAD}(\pi)$ quantifies policy reproducibility as:
\begin{equation}
\text{MAD}(\pi) = \text{Median}_\pi\left[\left\lvert\sum_{t=0}^{T-1}\gamma^{t}r_{t+1}(X_t,U_t) - \tilde{J}(\pi)\right\rvert\right]\,.
\end{equation}
In practice, this quantity is estimated from rolling out the policy over a number of episodesin the fixed environment.
Policies with high $\tilde{J}(\pi)$ but high dispersion may be unreliable in settings where consistent performance is critical, while those with slightly lower $\tilde{J}(\pi)$ but low dispersion may be preferable for reliable operation. Reporting both the performance measure and a dispersion metric enables a more comprehensive and practically relevant evaluation of RL policies in chemical process control.

In addition to performance and reproducibility measures, the optimality gap provides a crucial metric for evaluating reinforcement learning policies in chemical process control. The optimality gap, $\Delta(\pi_\phi)$, quantifies the difference between the performance of a learned policy $\pi_\phi$ and an optimal policy $\pi^*$:
\begin{equation}
    \Delta(\pi_\phi) = J(\pi^*) - J(\pi_\phi)
\end{equation}

where $J(\pi_\phi)$ and $J(\pi^*)$ represent the expected returns of the learned and optimal policies, respectively. This metric offers a normalized measure of policy performance across different environments and indicates the potential for improvement. In chemical process control, where true optimal policies may be unknown, upper bounds or benchmark policies such as nonlinear model predictive control with a perfect model {\sf (NMPC)} often serve as proxies for $\pi^*$.

Incorporating the optimality gap alongside performance measures like median return $\tilde{J}({\pi})$ and dispersion metrics such as MAD or standard deviation provides a comprehensive evaluation framework. This approach balances absolute performance, consistency, and relative optimality, enabling a more nuanced assessment of RL policies in the stochastic and complex domain of chemical processes. Such an evaluation is essential for developing robust and reliable control strategies that meet the stringent requirements of industrial chemical applications.

\subsection{Nonlinear MPC (Oracle)}
\label{sec:oracle}
To establish a performance benchmark for the RL algorithms, we implemented a nonlinear Model Predictive Controller {\sf (NMPC)} as an oracle. This controller leverages complete knowledge of the system dynamics, including perfect state estimation, to solve the optimal control problem at each time step. We define a normalization function for a vector $\textbf{z}$ as $\bar{\textbf{z}} = \text{diag}(\textbf{z}_{max} - \textbf{z}_{min})^{-1}(\textbf{z} - \textbf{z}_{min})$. The {\sf NMPC} is then formulated as a finite-horizon optimal control problem:

\begin{align*}
\min_{\textbf{u}_t, \ldots, \textbf{u}_{t+N-1}} J(\textbf{x}_t, \textbf{u}_t)& = \sum_{k=t+1}^{t+N} (\bar{\textbf{x}}_k - \bar{\textbf{x}}^*)^\top \textbf{Q} (\bar{\textbf{x}}_k - \bar{\textbf{x}}^*) + \sum_{k=t}^{t+N-1} (\bar{\textbf{u}}_k-\bar{\textbf{u}}_{k-1})^\top \textbf{R} (\bar{\textbf{u}}_k-\bar{\textbf{u}}_{k-1}) \\
 \text{s.t.} \quad 
 &\textbf{x}_t = \textbf{x}(t) \\
 & \textbf{x}_{k+1} = \textbf{f}(\textbf{x}_k, \textbf{u}_k, \textbf{w}_k), \quad k = t, \ldots, t+N-1 \\
&\textbf{g}(\textbf{x}_k) \leq \textbf{0}, \quad k = t, \ldots, t+N
\end{align*}

In this formulation, $\textbf{f}(\textbf{x}_k, \textbf{u}_k, \textbf{w}_k)$ represents a perfect model\footnote{To be explicit, here a perfect model means that the same realizations of uncertainty observed within the simulator are provided as data to define the nonlinear program. In this sense, the oracle can forecast the future perfectly.}  of the environment. The bar notation (e.g., $\bar{\cdot}_k$) indicates normalized variables, enabling balanced comparison between different variables, regardless of their physical units or magnitudes. The normalized state and setpoint vectors are denoted by $\bar{\textbf{x}}_k\in \mathbb{R}^{n_x}
$,  and $\bar{\textbf{x}}^* \in \mathbb{R}^{n_x}$, respectively; with $\bar{\textbf{u}}_k \in \mathbb{R}^{n_u}$ denoting the normalized input vector. The positive semi-definite weighting matrices for state and input costs are $\textbf{Q} \in \mathbb{R}^{n_x \times n_x}$ and $\textbf{R} \in \mathbb{R}^{n_u \times n_u}$, respectively. The vector-valued constraint function is represented by $\textbf{g}:\mathbb{R}^{n_x} \rightarrow \mathbb{R}^{n_g}
$. The subscript $k$ indicates the quantity at timestep $k$.

The oracle employs the {\sf do-mpc}~\cite{doMPC23} framework, which utilizes {\sf CasADi}~\cite{CasADi}, to formulate and {\sf IPOPT}~\cite{IPOPT06} to solve this optimal control problem. The {\sf NMPC} implementation uses orthogonal collocation to discretize the continuous-time dynamics, allowing for a high-accuracy representation of the system behavior between sampling instants. This approach is particularly beneficial for systems with fast dynamics or when using larger time steps.

The {\sf NMPC} problem is set up with a user-defined horizon length $N$ and control weighting $\textbf{R}$. The objective function includes terms for both state tracking error and control effort. The state tracking error is computed as the squared difference between each controlled variable's current state and setpoint. The control effort term is calculated using normalized input values to ensure consistent scaling across different input ranges.

This {\sf NMPC} implementation, with its perfect state estimation and high-accuracy collocation-based discretization, serves as a performance ceiling for the reinforcement learning algorithms. It represents near-optimal control given perfect system knowledge and disturbance prediction within the horizon. By comparing RL algorithm performance against this oracle, we can quantitatively assess their efficacy in approximating optimal control strategies across various process control scenarios.

\section{Software Implementation}
\label{sec:software_implementation}
\begin{figure}[ht!]
    \centering
    \includegraphics[width = \textwidth]{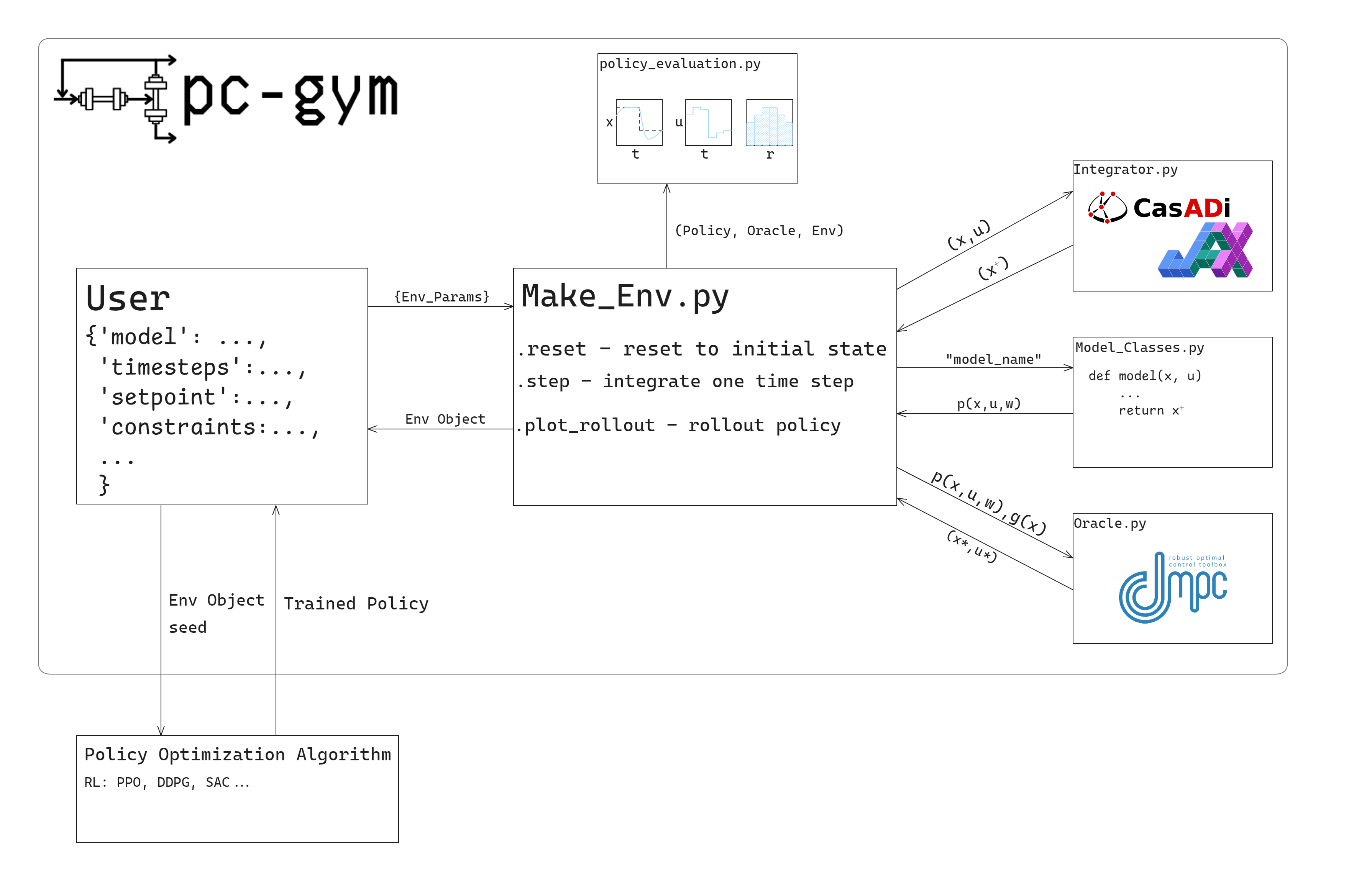}
    \caption{{\sf PC-Gym}'s Architecture}
    \label{fig:pc-gym}
\end{figure}
{\sf PC-Gym} presents a modular and extensible framework written in Python for simulating fixed control policies for complex dynamical systems. The architecture of {\sf PC-Gym} (Figure \ref{fig:pc-gym}) is designed to facilitate the integration of various components essential for reinforcement learning and optimal control tasks while adhering to established standards in the field.

Central to {\sf PC-Gym} is the \texttt{make\_env.py} module, which serves as the environment constructor. This module implements the Gymnasium \cite{gym} interface, a widely adopted standard for reinforcement learning environments. By following the Gymnasium format, {\sf PC-Gym} ensures compatibility with a broad range of existing RL algorithms and tools. The class accepts user-defined parameters as a dictionary, including model specifications, time steps, setpoints, and constraints.

In line with the Gymnasium standard, \texttt{make\_env.py} provides core methods such as \texttt{.reset()} for initializing the environment, and \texttt{.step()} for advancing the simulation. This standardization allows researchers and practitioners to seamlessly integrate {\sf PC-Gym} environments into their existing RL workflows and frameworks.

{\sf PC-Gym}'s modular design incorporates several key components that complement the Gymnasium-compatible environment. The \texttt{model\_classes.py} defines the dynamics of the controlled system. Numerical integration is handled by \texttt{integrator.py}, which utilizes either {\sf CASADI} \cite{CasADi} or {\sf JAX} \cite{jax2018github} (depending on the user's selection) for efficient computations. The oracle, as described in Section \ref{sec:oracle}, is implemented in \texttt{oracle.py} using the {\sf do-mpc}~\cite{doMPC23} toolkit, while \texttt{policy\_evaluation.py} manages policy assessment using the performance and dispersion metrics described in Section \ref{sec:policy_eval_background}. For more detailed information on {\sf PC-Gym}'s features and usage, readers are encouraged to refer to the documentation available at \href{https://maximilianb2.github.io/pc-gym/}{{\fontfamily{qcr}\selectfont
https://maximilianb2.github.io/pc-gym/}}.

\section{Control Problems}
\label{sec:controlproblems}
This section describes four environments included in {\sf PC-Gym}, indicating the types of dynamical systems included. The uncertainty in each control problem stems from measurement noise which given the perfect state estimator in the Oracle, results in no variation in the oracle's state and control, and therefore cost trajectories. At the end of this section, all environments are benchmarked with three RL algorithms (SAC, PPO, and DDPG) at a setpoint point tracking control problem. All RL algorithms are trained using 50,000 timesteps using the default hyperparameters from Stable Baselines 3~\cite{stable-baselines3} and the same seed. The choice of timestep budget was deemed large for this class of problems, allowing each algorithm to produce the highest performing policy for the same given seed. This policy is then fixed and evaluated on multiple instances of the environment. The following reward function follows the same notation as the Oracle's objective function as described in Section \ref{sec:oracle}:
\begin{equation}\label{eq:base_rew}
    r_{t+1}(\textbf{x}_{t}, \textbf{u}_t) = -\left((\bar{\textbf{x}}_{t+1} - \bar{\textbf{x}}^*)^\top \textbf{Q} (\bar{\textbf{x}}_{t+1} - \bar{\textbf{x}}^*) + (\bar{\textbf{u}}_{t}-\bar{\textbf{u}}_{t-1})^\top \textbf{R} (\bar{\textbf{u}}_t-\bar{\textbf{u}}_{t-1})\right) 
\end{equation}
where the state and control cost matrices, $\textbf{R}$ and \textbf{Q}, are set to equal the corresponding matrix in the oracle for each case study, and $\textbf{x}_{t+1}$ is obtained through forward simulation of the state space model given the current state and control pair, $(\textbf{x}_t, \textbf{u}_t)$ and realization of uncertain parameters.

\subsection{Continuously Stirred Tank Reactor} 

\label{sec:cstr}
The Continuous Stirred Tank Reactor (CSTR) system~\cite{Seborg2011, Ingham2007}, a complex and nonlinear dynamical system, is commonly studied in the Process Systems Engineering (PSE) literature. The system involves an irreversible chemical reaction converting species A to B, with cooling provided by water flowing through a coiled tube. Two ordinary differential equations describe the CSTR:

\begin{align}
  \frac{\mathrm{d}C_A}{\mathrm{d}t} &= \frac{q}{V}(C_{A_f} - C_A) - kC_Ae^{\frac{-E_A}{RT}}\\
  \frac{\mathrm{d}T}{\mathrm{d}t} &= \frac{q}{V}(T_f - T) -\frac{\Delta H_R}{\rho C_p}kC_Ae^{\frac{-E_A}{RT}} + \frac{UA}{\rho C_p V}(T_c - T)
\end{align}

The system features two state variables: concentration of species A, $C_A$, and reactor temperature, $T$. One input control variable, cooling water temperature, $T_c$. The system's nonlinearities are particularly evident in the exponential term containing the reactor temperature. The CSTR system presents significant challenges for control due to its nonlinear dynamics, strong coupling between state variables, and potential for instability and multiple steady states.




\subsection{Computational Implementation}
To demonstrate the {\sf PC-Gym} framework, we present the Python implementation used to construct the first case study. This example serves to illustrate how {\sf PC-Gym} significantly reduces the complexity associated with configuring advanced control environments, thereby enabling researchers to concentrate their efforts on the development and evaluation of control strategies, rather than expending resources on intricate implementation details. The following code segment demonstrates the concise methodology by which a CSTR control problem can be defined and initialized within the {\sf PC-Gym} framework.
\begin{minted}[
    frame=lines,
    framesep=2mm,
    baselinestretch=1.2,
    fontsize=\footnotesize,
]{python}
import numpy as np
from pcgym import make_env
from custom_reward import sp_track_reward 

T, tsim = 60, 25 # 60 timsteps for 25 minutes
x_star = {'Ca': [0.85] * (T // 3) + [0.9] * (T // 3) + [0.87] * (T // 3)} # Setpoints 

action_space = {
    'low': np.array([295]),
    'high': np.array([302])
}

observation_space = {
    'low': np.array([0.7, 300, 0.8]),
    'high': np.array([1, 350, 0.9])
}

# Dictionary which describes the process control problem
env_params = {
    'N': N,
    'tsim': tsim,
    'SP': x_star,
    'o_space': observation_space,
    'a_space': action_space,
    'x0': np.array([0.8, 330, 0.8]),
    'model': 'cstr',
    'noise_percentage': 0.001,
    'custom_reward': sp_track_reward # Define the reward as Equation 11
}

env = make_env(env_params) # environment object describing the process control problem
\end{minted}
The presented code above exemplifies the encapsulation of fundamental elements of the CSTR control problem within the {\sf PC-Gym} framework. By precisely defining critical parameters such as the number of timesteps ($N$), simulation duration (\texttt{tsim}), setpoints (\texttt{SP}), action and observation spaces, initial conditions (\texttt{x0}), and noise levels, researchers can expeditiously configure a realistic and challenging control scenario. We define a custom reward to mimic that shown in Equation \ref{eq:base_rew}. The final line of code instantiates the environment using the \texttt{make\_env} object. This singular operation demonstrates the simplicity with which {\sf PC-Gym} enables the creation of complex control problems, thereby establishing a robust foundation for subsequent control algorithm development and rigorous testing. This streamlined approach significantly reduces the barriers to entry for researchers in the field of process control.
\subsubsection{Results}
\label{sec:cstr_results}
\begin{figure}[ht!]
\centering
\includegraphics[clip, trim=0.24cm 0cm 0cm 0cm, width=1.00\textwidth]{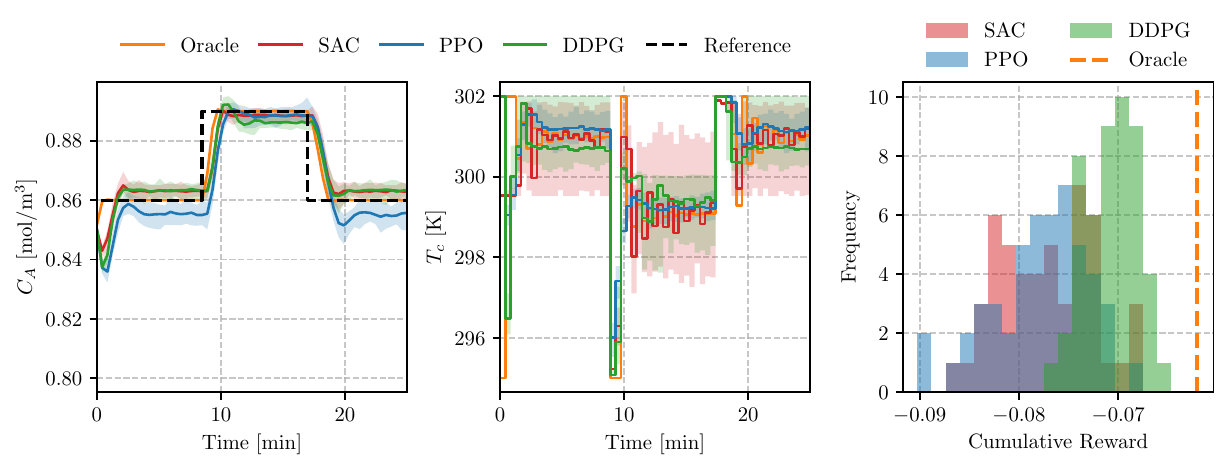}
\caption{Comparison of reinforcement learning algorithms (SAC, PPO, DDPG) and an oracle for control of the CSTR. Controlled variable response, manipulated variable trajectory, and cumulative reward distribution. Shaded areas show min-max ranges across simulations; solid lines represent median performance.}
\label{fig:cstr_results}
\end{figure}

The performance of three reinforcement learning algorithms was evaluated on the CSTR control problem and compared to the Oracle controller. The oracle used a prediction horizon of 17 steps, an identity $Q$ matrix, and a zero $R$ matrix. An episode was defined as 25 minutes with 60 timesteps. Figure \ref{fig:cstr_results} illustrates the control performance of these algorithms.

Figure \ref{fig:cstr_results} shows the controlled variable (C$_A$) response, the manipulated variable (T$_C$), and the distribution of reward for each algorithm. The oracle is represented as a single dashed line since measurement noise is the only source of uncertainty and the oracle has a perfect state estimator, therefor there is no variability in its performance. All three RL algorithms demonstrate the ability to track the setpoint changes, with SAC and DDPG showing the closest performance to the oracle. PPO also performs well but exhibits an offset for the first and third setpoints of $C_A$. SAC's control actions closely resemble the oracle's, with more aggressive transitions and minimal oscillations. DDPG's control actions are smoother than SAC and, with similar setpoint tracking performance, produce a higher reward. This reward histogram provides valuable insights into the performance and consistency of each algorithm. As expected, the oracle achieves the highest rewards with the narrowest distribution, indicating consistent near-optimal performance. DDPG demonstrates the best performance among the RL algorithms, with its reward distribution closest to the oracle. It achieves a high median reward albeit with a large spread similar to the other two reinforcement learning algorithms.

\subsection{Multistage Extraction Column}
The multistage extraction model \cite{Ingham2007} describes a counter-current liquid-gas extraction process with five stages. It captures the mass transfer of a solute between the liquid and gas phases in each stage. The model consists of ten ordinary differential equations (ODEs) representing the concentration dynamics of the solute in both phases for each stage.
The key equations for stage $n$, where $n \in {1, 2, 3, 4, 5}$, are:
\begin{align}
\frac{dC_{X_n}}{dt} &= \frac{1}{V_l}(L(C_{X_{n-1}} - C_{X_n}) - F_n) \\
\frac{dC_{Y_n}}{dt} &= \frac{1}{V_g}(G(C_{Y_{n+1}} - C_{Y_n}) + F_n)
\end{align}
where:
\begin{align}
C_{X_{n,eq}} &= \frac{C_{Y_{n,eq}}}{m} \\
F_n &= K_{la}(C_{X_n} - C_{X_{n,eq}})V_l
\end{align}
Here, $C_{X_n}$ and $C_{Y_n}$ are the solute concentrations in the liquid and gas phases, respectively, $L$ is the liquid flowrate, $G$ is the gas flowrate, $F_n$ is the mass transfer rate, and $C_{X_{n,eq}}$ is the equilibrium concentration.
The control variables are $L$ and $G$, while $C_{X_0}$ (feed concentration of liquid) and $C_{Y_6}$ (feed concentration of gas) are considered disturbances.
\begin{table}[h]
\centering
\caption{Multistage Extraction Model Parameters}
\begin{tabular}{@{}llc@{}}
\toprule
Parameter & Description & Value \\
\midrule
$V_l$ & Liquid volume in each stage (m$^3$) & 5.0 \\
$V_g$ & Gas volume in each stage (m$^3$) & 5.0 \\\
$m$ & Equilibrium constant (-) & 1.0 \\
$K_{la}$ & Mass transfer capacity constant (hr$^{-1}$) & 5.0 \\
$e$ & Equilibrium exponent & 2.0 \\
$C_{X_0}$ & Feed concentration of liquid (kg/m$^3$) & 0.60 \\
$C_{Y_6}$ & Feed concentration of gas (kg/m$^3$) & 0.050 \\
\bottomrule
\end{tabular}
\end{table}
This model represents a complex, multivariable system with nonlinear mass transfer dynamics. Nonlinearity is particularly evident in equilibrium relationships, which are governed by exponential terms. The counter-current nature of the process, coupled with the stage-wise mass transfer, makes this a challenging system for control applications. Typical control objectives might include maintaining desired output concentrations or optimizing extraction efficiency by manipulating the liquid and gas flow rates.

\subsubsection{Multistage Extraction Column Results}

\begin{figure}[ht!]
    \centering
    \includegraphics[width=1\linewidth]{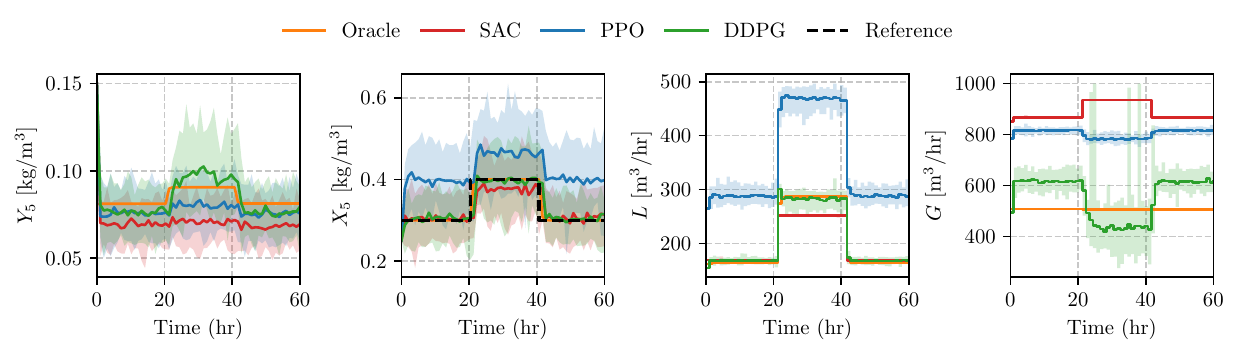}
    \caption{Comparison of reinforcement learning algorithms (SAC, PPO, DDPG) and an oracle for Multistage Extraction Column control. Controlled variable response, manipulated variable trajectory. Shaded areas show min-max ranges across simulations; solid lines represent median performance.}
    \label{fig:ME_states_controls}
\end{figure}

\begin{figure}[ht!]
    \centering
    \includegraphics[width=0.35\linewidth]{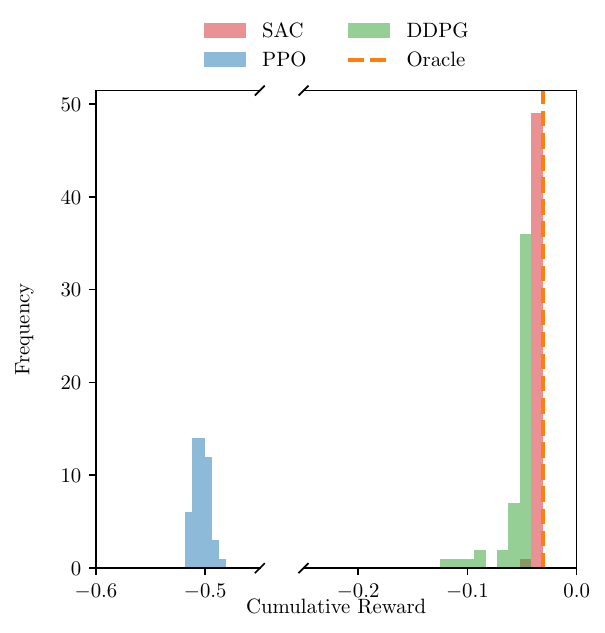}
    \caption{Cumulative reward distributions (100 repetitions) of reinforcement learning algorithms (SAC, PPO, DDPG) and an oracle for Multistage Extraction Column control.}
    \label{fig:ME_reward_dist}
\end{figure}

The performance of three reinforcement learning algorithms was evaluated on the multistage extraction column control problem of the bottom liquid phase $X_5$ and compared to the Oracle controller. The oracle used a prediction horizon of 20 steps, an identity $Q$ matrix, and a zero $R$ matrix. An episode was defined as 60 hours with 60 timesteps. Figure~\ref{fig:ME_states_controls} illustrates the control performance of these algorithms, showing both controlled variables ($Y_5$, $X_5$) and manipulated variables ($L$, $G$), while Figure~\ref{fig:ME_reward_dist} displays the distribution of cumulative rewards.

SAC demonstrated the best performance, closely matching the oracle in setpoint, although having different control profiles as SAC utilized the gas flow rate $G$. In contrast, the oracle maintained it at a constant value. It achieved the highest median reward, indicating near-optimal performance. DDPG performed well as it closely matched the setpoint; however, aggressive control of the gas flow rate on some of its runs resulted in a wider reward distribution than SAC and the oracle. PPO struggled with setpoint tracking, which had a significant offset throughout the episode, resulting in the poorest performance among the three algorithms. This was reflected in its wide reward distribution and lowest median reward.

\subsection{Crystallisation Reactor}
The crystallization model~\cite{de_moraes_modeling_2023} describes the population balance for potassium sulfate (K$_2$SO$_4$) crystallization using the method of moments. It is a highly nonlinear process represented by a system of ordinary differential equations (ODEs) that capture the evolution of the first four moments of the crystal size distribution ($\mu_0$, $\mu_1$, $\mu_2$, $\mu_3$) and the solute concentration ($c$). The model incorporates nucleation and growth kinetics, which are influenced by supersaturation and temperature.
The key equations are:
\begin{align}
\frac{d\mu_0}{dt} &= B_0 \\
\frac{d\mu_1}{dt} &= G_{\infty} (a\mu_0 + b\mu_1 \times 10^{-4}) \times 10^4 \\
\frac{d\mu_2}{dt} &= 2G_{\infty} (a\mu_1 \times 10^{-4} + b\mu_2 \times 10^{-8}) \times 10^8 \\
\frac{d\mu_3}{dt} &= 3G_{\infty} (a\mu_2 \times 10^{-8} + b\mu_3 \times 10^{-12}) \times 10^{12} \\
\frac{dc}{dt} &= -0.5\rho\alpha G_{\infty} (a\mu_2 \times 10^{-8} + b\mu_3 \times 10^{-12})
\end{align}
where $B_0$ is the nucleation rate and $G_{\infty}$ is the crystal growth rate. These rates depend on the supersaturation $S$ and temperature $T$, which is the control variable:
\begin{align}
C_{eq} &= -686.2686 + 3.579165(T+273.15) - 0.00292874(T+273.15)^2 \\
S &= c \times 10^3 - C_{eq} \\
B_0 &= k_a \exp\left(\frac{k_b}{T+273.15}\right) (S^2)^{k_c/2} (\mu_3^2)^{k_d/2} \\
G_{\infty} &= k_g \exp\left(\frac{k_1}{T+273.15}\right) (S^2)^{k_2/2}
\end{align}
The model parameters are given in the following table:
\begin{table}[h]
\centering
\caption{Crystallization Model Parameters}
\begin{tabular}{@{}ccc@{}}
\toprule
Parameter & Description & Value \\
\midrule
$k_a$ & Nucleation rate constant & 0.92 \\
$k_b$ & Nucleation temperature dependency & -6800 \\
$k_c$ & Nucleation supersaturation exponent & 0.92 \\
$k_d$ & Nucleation crystal content exponent & 1.3 \\
$k_g$ & Growth rate constant & 48 \\
$k_1$ & Growth rate temperature dependency & -4900 \\
$k_2$ & Growth rate supersaturation exponent & 1.9 \\
$a$ & Size-dependent growth parameter & 0.51 \\
$b$ & Size-dependent growth parameter & 7.3 \\
$\alpha$ & Volumetric shape factor & 7.5 \\
$\rho$ & Crystal density (g/cm$^3$) & 2.7 \\
\bottomrule
\end{tabular}
\end{table}
This model captures the complex crystallization dynamics, including the effects of temperature on nucleation and growth rates, making it a challenging system for control applications. The control objective typically involves manipulating the temperature to achieve desired crystal size distribution characteristics or solute concentration levels.
\subsubsection{Crystallization Reactor Results}

\begin{figure}[ht!]
\centering
\includegraphics[clip, trim=0.24cm 0cm 0cm 0cm, width=1.00\textwidth]{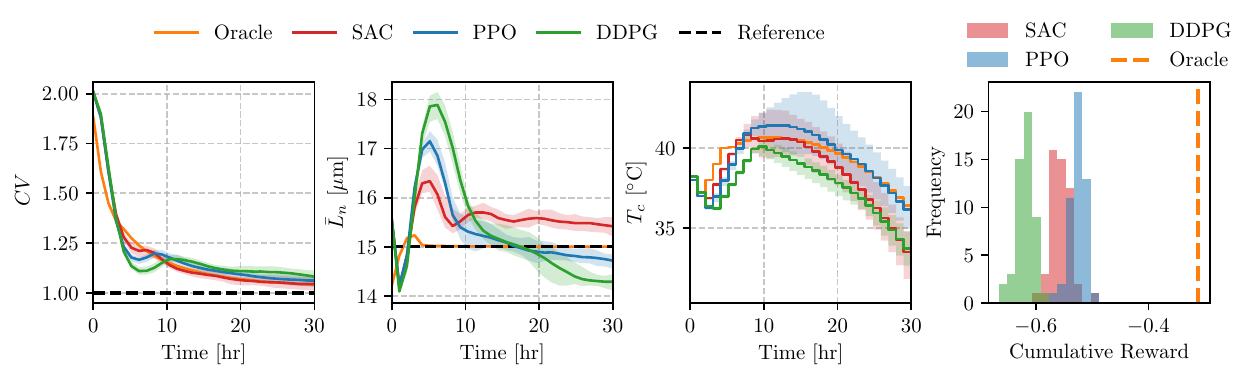}
\caption{Comparison of reinforcement learning algorithms (SAC, PPO, DDPG) and an oracle for Crystalization Reactor control. Controlled variable response, manipulated variable trajectory, and cumulative reward distribution. Shaded areas show ranges across simulations; solid lines represent median performance.}
\label{fig:crystalisation_results}
\end{figure}
The performance of three reinforcement learning algorithms was evaluated on the crystalization control problem and compared to the oracle controller. The oracle used a prediction horizon of 2 steps: an identity $Q$ matrix and a zero $R$ matrix. An episode was defined as 30 hours with 30 timesteps. Figure~\ref{fig:crystalisation_results} illustrates the control performance of these algorithms, showing both controlled variables ($CV$, $\bar{L}_n$) and manipulated variable ($T_c$), this figure also displays the distribution of cumulative rewards.

All three reinforcement learning algorithms perform poorly in tracking $\bar{L}_n$ with a significant overshoot in the first 10 hours compared to the oracle. Out of the three reinforcement learning algorithms, PPO performs the best. This is shown by a higher median reward in the cumulative reward distribution, also shown in Figure~\ref{fig:crystalisation_results}.
\subsection{Four-Tank System}
The four-tank system~\cite{Johansson2000} is a multivariable process consisting of four interconnected water tanks. This system is often used as a benchmark for control systems due to its nonlinear dynamics and the coupling between inputs and outputs. The model describes the change in water levels in each tank based on the inflows and outflows.
The key equations for each tank are:
\begin{align}
\frac{dh_1}{dt} &= -\frac{a_1}{A_1}\sqrt{2g_ah_1} + \frac{a_3}{A_1}\sqrt{2g_ah_3} + \frac{\gamma_1 k_1}{A_1}v_1 \\
\frac{dh_2}{dt} &= -\frac{a_2}{A_2}\sqrt{2g_ah_2} + \frac{a_4}{A_2}\sqrt{2g_ah_4} + \frac{\gamma_2 k_2}{A_2}v_2 \\
\frac{dh_3}{dt} &= -\frac{a_3}{A_3}\sqrt{2g_ah_3} + \frac{(1-\gamma_2)k_2}{A_3}v_2 \\
\frac{dh_4}{dt} &= -\frac{a_4}{A_4}\sqrt{2g_ah_4} + \frac{(1-\gamma_1)k_1}{A_4}v_1
\end{align}
Here, $h_i$ represents the water level in tank $i$, $v_1$, and $v_2$ are the input voltages to the pumps, and the other parameters are described in the table below.
\begin{table}[h]
\centering
\caption{Four-Tank System Model Parameters}
\begin{tabular}{@{}llc@{}}
\toprule
Parameter & Description & Value \\
\midrule
$g_a$ & Acceleration due to gravity (m/s$^2$) & 9.8 \\
$\gamma_1$ & Fraction bypassed by valve to tank 1 (-) & 0.20 \\
$\gamma_2$ & Fraction bypassed by valve to tank 2 (-) & 0.20 \\
$k_1$ & 1st pump gain (m$^3$/Volts·s) & 8.5 × 10$^{-4}$ \\
$k_2$ & 2nd pump gain (m$^3$/Volts·s) & 9.5 × 10$^{-4}$ \\
$a_1$ & Cross-sectional area of outlet of tank 1 (m$^2$) & 3.5 × 10$^{-3}$ \\
$a_2$ & Cross-sectional area of outlet of tank 2 (m$^2$) & 3.0 × 10$^{-3}$ \\
$a_3$ & Cross-sectional area of outlet of tank 3 (m$^2$) & 2.0 × 10$^{-3}$ \\
$a_4$ & Cross-sectional area of outlet of tank 4 (m$^2$) & 2.5 × 10$^{-3}$ \\
$A_1$ & Cross-sectional area of tank 1 (m$^2$) & 1.0 \\
$A_2$ & Cross-sectional area of tank 2 (m$^2$) & 1.0 \\
$A_3$ & Cross-sectional area of tank 3 (m$^2$) & 1.0 \\
$A_4$ & Cross-sectional area of tank 4 (m$^2$) & 1.0 \\
\bottomrule
\end{tabular}
\end{table}
This system is characterized by its multivariable nature and the coupling between inputs and outputs. The nonlinearity is evident in the square root terms, representing each tank's outflow rates. The system's behavior can be adjusted by changing the valve positions ($\gamma_1$ and $\gamma_2$), allowing minimum and non-minimum phase configurations.
The control objective typically involves regulating the water levels in the lower tanks (h$_1$ and h$_2$) by manipulating the input voltages to the pumps (v$_1$ and v$_2$). This presents a challenging control problem due to the system's nonlinearity, coupling, and potential for non-minimum phase behavior.

\subsubsection{Four-Tank System Results}
\begin{figure}[ht!]
    \centering
    \includegraphics[width=1\linewidth]{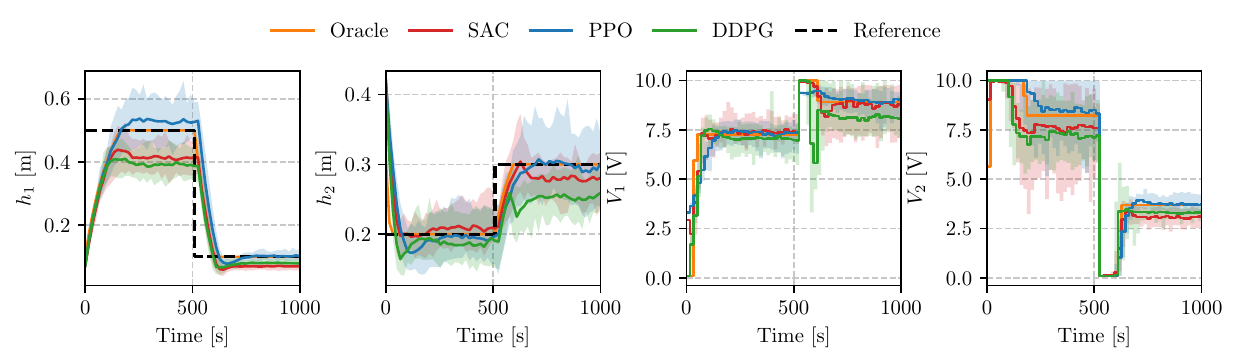}
    \caption{Comparison of reinforcement learning algorithms (SAC, PPO, DDPG) and an oracle for Four Tank System control. Controlled variable responses and manipulated variable trajectories. Shaded areas show min-max ranges across simulations; solid lines represent median performance.}
    \label{fig:4tank_states_controls}
\end{figure}

\begin{figure}[ht!]
    \centering
    \includegraphics[width=0.35\linewidth]{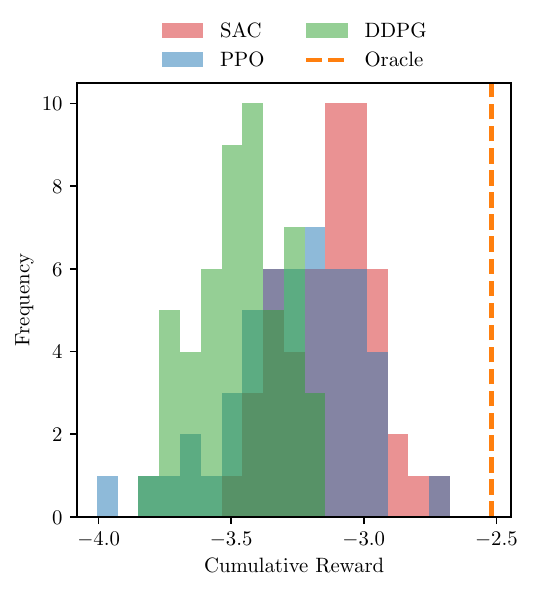}
    \caption{Cumulative reward distributions (20 repetitions) of reinforcement learning algorithms (SAC, PPO, DDPG) and an oracle for Multistage Extraction Column control.}
    \label{fig:4tank_reward_dist}
\end{figure}

The performance of three reinforcement learning algorithms was evaluated on the Four Tank system control
problem and compared to the Oracle controller. The oracle MPC used a prediction horizon of 17 steps, an identity Q
matrix, and a zero R matrix. An episode was defined as 1000 seconds with 60 timesteps. Figure \ref{fig:4tank_states_controls} illustrates the control
performance of these algorithms, showing both controlled variables ($h_1$, $h_2$) and manipulated variables ($V_1$, $V_2$), while
Figure \ref{fig:4tank_reward_dist} displays the distribution of cumulative rewards.

SAC achieved the highest median reward with a narrow distribution, indicating consistent performance albeit with a significant offset for the level of Tank One. PPO showed a lower median reward but with a wide distribution, occasionally reaching SAC's performance levels. DDPG presented the lowest median reward due to significant offset for the first setpoint for $h_1$ and the second setpoint for $h_2$. 

\subsection{Overview of Case Studies}
The optimality gap, as previously discussed in Section \ref{sec:policy_eval_background}, holds significant importance in reinforcement learning for several reasons. Primarily, it offers a scalar measure that quantifies how far a learned policy is from optimality, enabling direct comparisons between different algorithms. For our study in {\sf PC-Gym} environments, we calculate the empirical optimality gap with normalized cumulative rewards, and with the optimal policy approximated by the oracle $\pi^* \approx \pi_o$:

\begin{equation}
    \hat{\Delta}(\pi_\phi) =  \tilde{J}(\pi_o) -   \tilde{J}(\pi_\phi)
\end{equation}
where $M$ is the number of episodes, and $\tilde{J}(\pi)$ is the median return under policy $\pi$ where $\pi_o$ represents the oracle and $\pi_\phi$ represents an RL policy either optimized with SAC, PPO, or DDPG. It should be noted that $\tilde{J}(\cdot)$ uses normalized return values to allow for comparison between case studies.

\begin{table}[]
\centering

\caption{Optimality Gap of SAC, DDPG, and PPO in the {\sf PC-Gym} Environments}
\label{tab:opt_gap}
\begin{tabular}{@{}cccc@{}}
\toprule
\multirow{2}{*}{Environment} & \multicolumn{3}{c}{Optimality Gap} \\
                             & DDPG       & SAC       & PPO       \\ \midrule
CSTR                         &  \textbf{0.008}       & 0.014       & 0.0157       \\
Multistage Extraction        &   0.019       & \textbf{0.002}         &0.474         \\
Crystallization              & 0.304           & 0.246          &      \textbf{0.211}     \\
Four Tank System             &0.950           &    \textbf{0.579}       & 0.743          \\
 \bottomrule
\end{tabular}
\end{table}
Comparing the optimality gaps (Table~\ref{tab:opt_gap}) across the four {\sf PC-Gym} environments reveals varying performance among DDPG, SAC, and PPO. DDPG demonstrates superior performance in the CSTR environment with the lowest gap of 0.008. SAC excels in the Multistage Extraction and Four Tank System, achieving gaps of 0.002 and 0.579, respectively. PPO performs best in the Crystallization environment with a gap of 0.211. SAC shows consistent performance across environments, while DDPG and PPO exhibit more variability, suggesting that algorithm selection should be tailored to the specific control task.

\begin{table}[]
\centering
\caption{Median Absolute Deviation of SAC, DDPG and PPO for the {\sf PC-Gym} Environments}
\label{tab:MAD}
\begin{tabular}{@{}cccc@{}}
\toprule
\multicolumn{1}{c}{\multirow{2}{*}{Policy}} & \multicolumn{3}{l}{Median Absolute Deviation} \\
\multicolumn{1}{c}{}                        & DDPG          & SAC          & PPO          \\ \midrule
CSTR                                        &   \textbf{0.0013 }           &0.0035              &0.0022              \\
Multistage Extraction                       &  0.0021             & \textbf{0.0001 }             &  0.0053            \\
Crystallization                             &    0.009           & 0.013              &\textbf{0.007}              \\
Four Tank                                   &0.124               &\textbf{0.115 }             &0.145              \\ \bottomrule
\end{tabular}
\end{table}
The MAD (Table~\ref{tab:MAD}) provides insight into the consistency and robustness of each algorithm's performance across multiple runs. Lower MAD values indicate more stable performance. DDPG shows the most consistent performance in the CSTR environment, while SAC demonstrates high stability in the Multistage Extraction process. PPO exhibits the lowest variability in the Crystallization environment. The Four Tank system presents challenges for all algorithms, with SAC showing slightly better consistency. These results highlight the importance of considering both performance and stability when selecting an RL algorithm for process control tasks.

To conclude this case study, our analysis of the optimality gap and median absolute deviation across different {\sf PC-Gym} environments reveals that no single algorithm consistently outperforms the others in all scenarios. The choice of the most suitable reinforcement learning algorithm depends on the specific characteristics of the process control task at hand. SAC demonstrates a good balance between performance and consistency across various environments, making it a strong general-purpose choice. However, DDPG and PPO show superior performance in specific scenarios, indicating that they may be preferable for certain types of control problems. Furthermore, this case study evaluates fixed policies with a large evaluation budget for training therefore it cannot be said how these algorithms may perform in a sample-constraint and online learning setting. These findings underscore the importance of careful algorithm development required for process control applications, where both performance and stability are crucial factors.

\section{Process Disturbances}\label{sec:disturbance}

In chemical process control, disturbances significantly impact system behavior and performance. They can originate from various sources, such as fluctuations in feed composition, environmental changes, or equipment malfunctions. These disturbances can lead to deviations from desired operating conditions, potentially affecting product quality, energy efficiency, and safety. Consequently, the ability to effectively handle and mitigate the impact of disturbances is a key aspect of robust process control.
\subsection{Disturbance Generation}
To demonstrate the flexibility in disturbance generation of {\sf PC-Gym}, the CSTR example is used from Section \ref{sec:cstr} with a disturbance in the inlet temperature. To train the agent, an environment was created where an uncertain step change was applied to the inlet temperature after 20 timesteps had passed, then at 40 timesteps the inlet temperature reverted to the original value whilst keeping the reactor temperature at a constant setpoint of 340 K. These inlet temperatures were bounded between 330 $K$ and 370 $K$. Since SAC and DDPG performed well on the CSTR example in Section \ref{sec:cstr_results} these two algorithms were used to train agents with the default hyperparameters from Stable Baselines 3 \cite{stable-baselines3}. 
\begin{equation}
T_{in} =
\begin{cases}
350\text{ K} & \text{if } i < 20 \text{ or } i > 40 \\
\mathcal{U}(330\text{ K}, 370\text{ K}) & \text{if } 20 \leq i \leq 40
\end{cases}
\end{equation}
where $\mathcal{U}(L, U)$ represents a 1D uniform distribution with a lower bound of $L$ and upper bound of $U$. Both SAC and DDPG were used to train an agent on this distribution of disturbances for 50,000 timesteps each. One disturbance from the training distribution and one outside of each algorithm were chosen to test each algorithm.
\subsection{Computational Implementation}
To generate the disturbances for this experiment in {\sf PC-Gym} the user will have to define two dictionaries. The first is the disturbance bounds which define the disturbance space that completes the observation space. his dictionary specifies the lower and upper bounds for each disturbance variable. The second is the disturbance dictionary, which employs disturbance variable names as keys and their corresponding time-indexed values as entries. This structure allows users to specify the evolution of disturbances over time. These disturbances are treated as additional control inputs to the model, which either take the model's default value or the user's disturbance value. This retains flexibility in the definition of the disturbance. For this experiment, the computational implementation of the environment creation (parameters not unique to disturbance generation were omitted for readability) is as follows: 
\begin{minted}[
    frame=lines,
    framesep=2mm,
    baselinestretch=1.2,
    fontsize=\footnotesize,
]{python}
import numpy as np
from pcgym import make_env

# Define disturbance bounds
disturbance_bounds = {
    'low': np.array([330]),
    'high': np.array([370])
}


# Environment parameters
env_params_template = {
                        # ...other parameters ...
                        'disturbance_bounds': disturbance_bounds
                        # Disturbance generated from uniform distribution
                        'disturbance': {'Ti': [350] * (T // 3) 
                                            + [np.random.uniform(330, 370)] * (T // 3) 
                                            + [350] * (T // 3)}
                        # ... other parameters ...
                        }

env = make_env(env_params) # environment object describing the process control problem
\end{minted}

\subsection{Results}
\begin{figure}[ht!]
    \centering
    \includegraphics[width = 1 \textwidth]{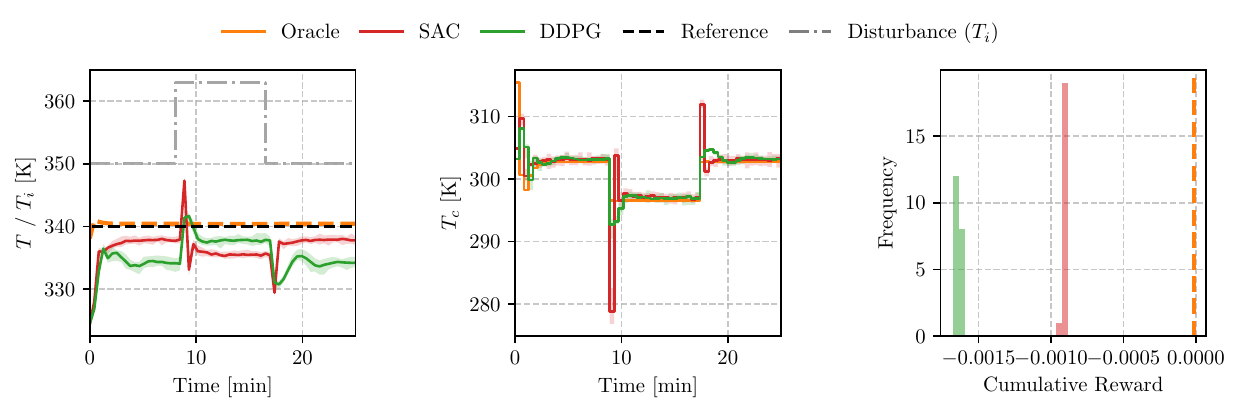}
    \caption{Controlled variable and disturbance trajectories, the manipulated variable trajectories, and cumulative reward distribution for a step change of 350 K to 363 K to the inlet temperature.}
    \label{fig:1stdist}
\end{figure}

The agents are first tested on one sample from the training distribution of disturbance, the resulting controlled variable and manipulated variable trajectories for both RL agents and the oracle are shown in Figure \ref{fig:1stdist}. The disturbance shown in this figure is a step change of 350 K to 363 K of the inlet temperature. Both the RL agents exhibit significant offset throughout the simulation compared to the oracle, which can track the setpoint with negligible error. However, when the disturbance occurs, the agent trained with SAC performs aggressive control actions to attempt to reject the disturbance, whereas the agent trained with DDPG performs more conservative control, which results in less effective disturbance rejection. This can be quantified with the optimality gap, where over 10 samples of the training distribution, the agent trained with SAC has a mean normalized optimality gap of 76.58 compared to 138.97 for the agent trained with DDPG. \\

\begin{figure}[ht!]
    \centering
    \includegraphics[width = 1\textwidth]{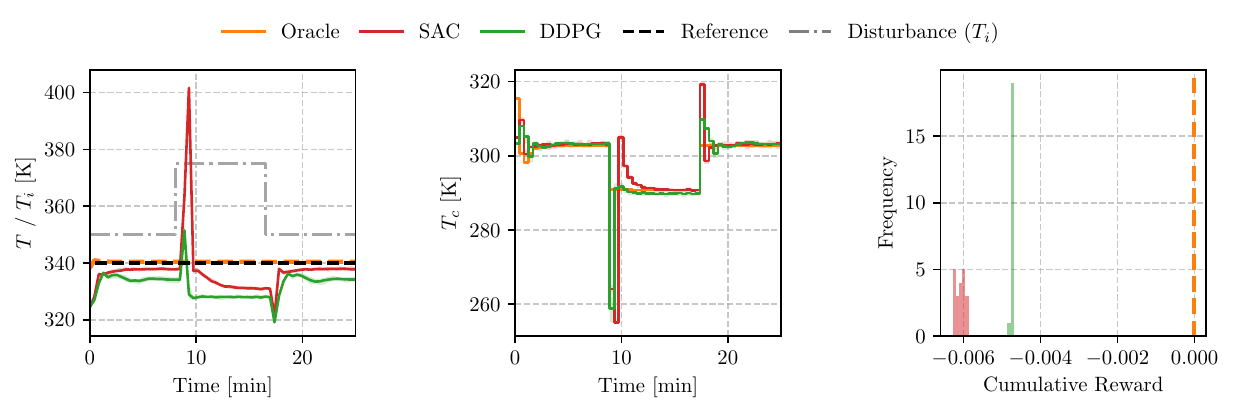}
    \caption{Controlled variable and disturbance trajectories, the manipulated variables trajectories, and cumulative reward distribution for a step change of 350 K to 375 K to the inlet temperature.}
    \label{fig:altdist}
\end{figure}

Both agents were then tested on a disturbance from outside of the training distribution, and this was selected as a step change to 375 K. The resulting controlled and manipulated variable trajectories, along with the distribution of rewards for both agents and the oracle, are shown in Figure \ref{fig:altdist}. The policy learned using SAC, which was effective within the training distribution, now results in an overly aggressive control action, which leads to a large spike in reactor temperature. On the other hand, the policy trained with DDPG rejects the disturbance more effectively, given its conservative control actions. The ranking of agents in terms of optimality gap swaps position with the agent trained with DDPG performing better with a normalized optimality gap on this disturbance of 408.3 compared to 525.08 for the agent trained with SAC.\\

The demonstration of {\sf PC-Gym}'s disturbance generation capabilities in the CSTR example highlights its flexibility and utility in training robust control agents. By allowing users to define time-indexed disturbance values, {\sf PC-Gym} enables the creation of realistic and challenging scenarios for reinforcement learning in process control. The ability to easily implement and evaluate various disturbance patterns in {\sf PC-Gym} will accelerate the development of adaptive and robust process control using reinforcement learning techniques.

\newpage
\section{Constraint Handling}
\label{sec:constraints}
In chemical process control, operating within safe and efficient boundaries is crucial. Our library implements a flexible constraint-handling mechanism to model these real-world limitations effectively. This section demonstrates constraint's impact on RL in chemical process environments. Here, the focus is to demonstrate the flexibility of {\sf PC-Gym} in imposing state 
constraints. Many constrained-RL algorithms have been developed, which could be deployed \cite{zhang2020first,petsagkourakis2022chance, mowbray2022safe}. In the following, we demonstrate a simple reward shaping strategy for the purpose of demonstration.

\subsection{Reward Shaping for Constraint Satisfaction}
To encourage constraint satisfaction without hard termination, we implement a custom reward function that incorporates a significant penalty for violations:
\begin{equation}\label{eq:con_reward}
r_{con}(\textbf{x}_t, \textbf{u}_t) = r(\textbf{x}_t, \textbf{u}_t) - \lambda \sum_{i=0}^{n_g} \max(0, g_i(\textbf{x}_t, \textbf{u}_t))
\end{equation}
where $r$ is the base reward function as described in Equation \ref{eq:base_rew}, $\lambda > 0$ is a penalty coefficient, and the sum term represents the cumulative constraint violation. This approach effectively shapes the reward landscape, guiding the RL agent towards constraint-satisfying policies while maintaining exploration in the full state space. This is not the only way to handle constraints within the reinforcement learning framework, for further reading we refer to works within the safe RL community. This approach was used to demonstrate the constraint-creation capabilities of {\sf PC-Gym}.

\subsection{Computational Implementation}
The creation of the environment used in the constraint showcase is shown below. First, the constraints and their type are defined as dictionaries where the key is the constrained state, and the value is either a list of constraints and types. Then these are passed to the environment parameter dictionary and subsequently to the \texttt{make\_env} class. The constraint violation information for the current timestep can be accessed from the tuple returned by the \texttt{.step} method of the environment object. 
\begin{minted}[
    frame=lines,
    framesep=2mm,
    baselinestretch=1.2,
    fontsize=\footnotesize,
]{python}
import numpy as np
from pcgym import make_env
from custom_reward import con_reward

# Constraints and their type are defined
cons = {'T':[327,321]}
cons_type = {'T':['<=','>=']}

env_params_con = {
    ...
    'constraints': cons, # Constraints are added to the environment 
    'cons_type': cons_type, # Constraint types are added to the environment
    ...
    }
    
env_con = make_env(env_params_con) # Environment object describing the process control problem
\end{minted}
\subsection{Experimental Results}
\begin{figure}[htbp]
\centering
\includegraphics[width = \textwidth]{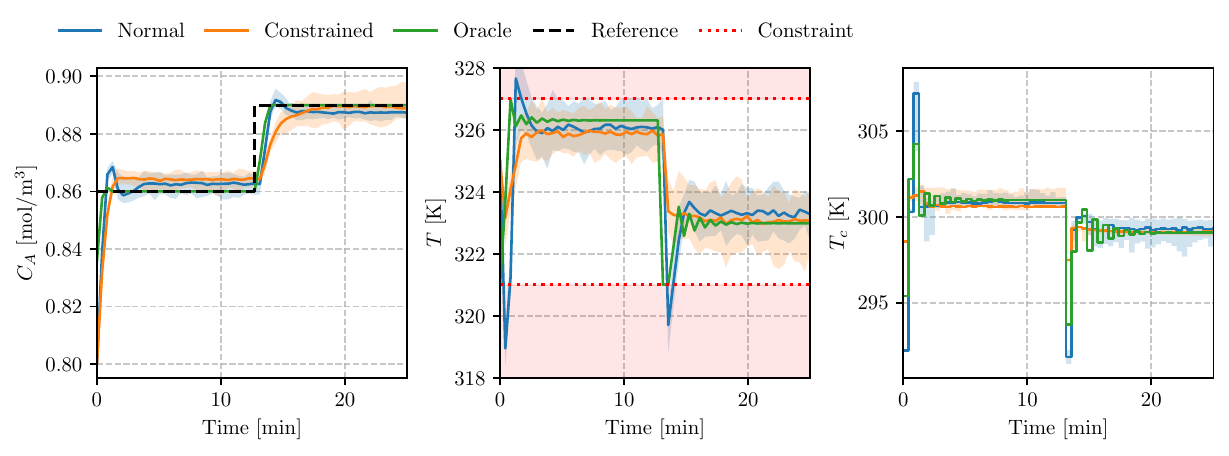}
\caption{Rollout comparison of constrained and unconstrained DDPG agents over 50 repetitions}
\label{fig:rollout_comparison}
\end{figure}

We conducted experiments using the CSTR environment introduced in Section \ref{sec:cstr} to evaluate our constraint mechanism. Two DDPG agents were trained: one with reward shaped according to Equation \ref{eq:con_reward} with $\lambda = 1$ to equally weight constraint satisfaction and setpoint tracking, and one with the base reward function as shown in Equation \ref{eq:base_rew}. The environment simulated a typical CSTR operation with setpoint changes in reactant concentration and included temperature constraints of the form:
\begin{equation}
321\hspace{2pt}\text{K} \leq T \leq 327\hspace{2pt}\text{K}
\end{equation}
Figure \ref{fig:rollout_comparison} shows the rollout results for both agents, focusing on the temperature state variable and control actions.
The constrained agent successfully learned to maintain the temperature within the specified bounds, as evident from the temperature trajectory in Figure \ref{fig:rollout_comparison}. In contrast, the base agent violated these limits in pursuit of optimal setpoint tracking for the concentration (Figure \ref{fig:rollout_comparison}), especially in the overshoot of both setpoints.
To quantify constraint satisfaction, we leverage the empirical cumulative probability of constraint violation:
\begin{equation}
P_{violation}(\pi) = \frac{1}{N} \sum_{i=1}^N I(g(\tau_i) > 0)
\end{equation}
where $N$ is the total number of episodes, $I$ is the indicator function that evaluates the entire trajectory of an episode, and $\tau_i = {(\textbf{x}_t, \textbf{u}_t)}_{t=0}^T$ represents the complete state-action trajectory of the $i$-th episode from time step 0 to $T$. The violation metric results for the experiment are shown in Table~\ref{tab:con_violation}.

\begin{table}[ht!]
\caption{Empirical constraint violation probability for the base reward, shaped reward, and the Oracle}
\label{tab:con_violation}
\centering
\begin{tabular}{@{}cc@{}}
\toprule
Policy      & Empirical Violation Probability ($V$) \\ \midrule
Base reward      & 0.0727                              \\
Shaped reward & 0.0003                              \\
Oracle      & 0.0000                              \\ \bottomrule
\end{tabular}
\end{table}

From the violation metric, the shaping of the reward significantly reduces the probability compared to the base case. However, there is still a substantial optimality gap to the oracle of 0.016. This shows that the RL agent could operate less conservatively and improve setpoint tracking whilst satisfying the constraints. These results demonstrate that {\sf PC-Gym}'s constraint handling mechanism provides a robust foundation for studying and implementing constrained RL in chemical process control. It offers the flexibility necessary for realistic scenarios while enabling systematic comparison of different safe and efficient learning approaches.
\section{Reward Functions}\label{sec:rewardfunc}                           
In process control applications, the optimal control strategy often depends on the specific goals and limitations of the system. For example, in a CSTR control task, the reward function might prioritize maintaining the reactor concentration within a certain range, minimizing energy consumption, or reducing the variability of the controlled variable. By allowing users to define their reward functions, {\sf PC-Gym} provides the flexibility to model the requirements of the process control problem. To showcase this flexibility, we will demonstrate three ways to design reward functions for a setpoint tracking task: absolute error, square error, and a sparse reward formulation. The first formulation, negative absolute error (\ref{eq:l1_norm}), will reward the agent for being close to setpoint ($\textbf{x}^*$) and is defined by:

\begin{equation}\label{eq:l1_norm}
    r_{t+1}(\textbf{x}_t, \textbf{u}_t)  = -\lvert| \bar{\textbf{x}}^* - \bar{\textbf{x}}_{t+1}\rvert|_1
\end{equation}

The negative squared error (\ref{eq:squared_error}) will also reward the agent for being close to the setpoint. However, it will penalize the agent more heavily than the absolute error as it moves away from the set point.

\begin{equation}\label{eq:squared_error}
    r_{t+1}(\textbf{x}_t, \textbf{u}_t) = -\lvert|\bar{\textbf{x}}^* - \bar{\textbf{x}}_{t+1}\rvert|_2
\end{equation}

Furthermore, a sparse reward (\ref{eq:sparse}) can be defined where a reward is only given when the squared setpoint error is smaller than a threshold value $\epsilon$.
\begin{equation}\label{eq:sparse}
r_{t+1}(\textbf{x}_t, \textbf{u}_t) = 
\begin{cases}
1, & \text{if } -\lvert|\bar{\textbf{x}}^* - \bar{\textbf{x}}_{t+1}\rvert|_2 < \epsilon \\
0, & \text{otherwise}\,.
\end{cases}
\end{equation}
In all cases, the next state $\bar{\textbf{x}}_{t+1}$ is obtained through forward simulation of the state space model, given a realization of uncertain parameters and the current state-control pair.

\subsection{Computational Implementation}
The following Python code was used to implement the training of three different reward functions within {\sf PC-Gym}. First, a list of reward functions is defined which is then looped over to create an environment and subsequent policy with each reward function. The code unique to the creation of this experiment is shown below:
\begin{minted}[
    frame=lines,
    framesep=2mm,
    baselinestretch=1.2,
    fontsize=\footnotesize,
]{python}

from pcgym import make_env
from custom_reward import r_sparse, r_squared, r_abs

# Define a list of reward functions
r_func = [r_squared,r_sparse, r_abs]

# loop through the custom reward functions
for r_func_i in r_func: 
  env_params = {
                # ... other parameters ... 
                'custom_reward': r_func_i 
                # ... other parameters ...
                }
    
    env = make_env(env_params)

\end{minted}

\subsection{Results}
\begin{figure}[htbp]
    \centering
    \includegraphics[width=\textwidth]{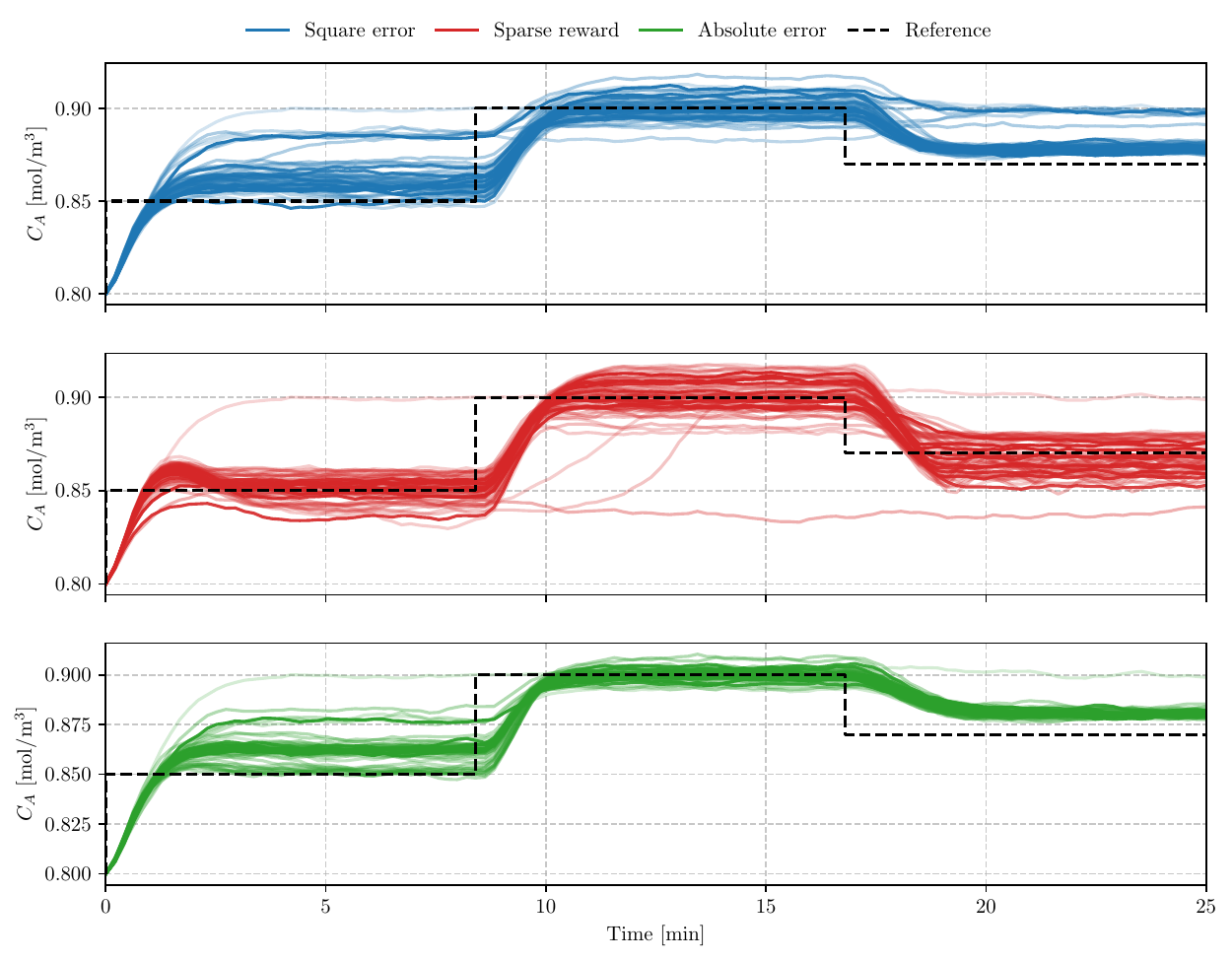}
    \caption{Comparison of learning progress for different reward functions in DDPG. The plot shows the concentration of reactant A ($C_A$) over time for three reward strategies: square error (top), sparse reward (middle), and absolute error (bottom). The dashed black line represents the reference trajectory. Lighter colors show later iterations.}
    \label{fig:reward_learning}
\end{figure}

These three reward functions are implemented with {\sf PC-Gym}'s custom reward functionality on the CSTR environment with $\epsilon$ set to 0.003 for the sparse reward function. Three DDPG reinforcement learning agents are all trained for 15k timesteps with the default hyperparameters from Stable Baselines 3 \cite{stable-baselines3}. The trajectories sampled from the current policy every 500 timesteps of training are shown in Figure \ref{fig:reward_learning}.

Figure \ref{fig:reward_learning} shows that different reward functions designed for the same task, setpoint tracking, can produce significantly different policies. The sparse reward signal can make it more difficult for the agent to learn, especially in the early stages of training when the agent rarely receives a non-zero reward, and it cannot further improve the setpoint error below the threshold value provided. The square error leads to slower learning when the agent is close to the setpoint. In contrast, the absolute error provides a consistent reward signal proportional to the setpoint error, allowing it to track the setpoint well.
\newpage
\section{Conclusion and future works}
\label{sec:conclusion}
{\sf PC-Gym} provides a comprehensive framework for applying reinforcement learning to chemical process control problems. Offering discrete-time environments with nonlinear dynamics, flexible constraint handling, disturbance generation, and customizable reward functions enables the evaluation of RL algorithms against traditional control methods. The framework's modular design makes it straightforward for users to implement custom environments, allowing researchers to easily extend {\sf PC-Gym}'s capabilities to their specific process control scenarios while maintaining compatibility with the core framework features.
Our benchmarking results across various process scenarios demonstrate {\sf PC-Gym}'s effectiveness in assessing RL approaches for complex control tasks. This framework bridges the gap between theoretical RL advancements and practical applications in industrial process control, paving the way for more improved control strategies. {\sf PC-Gym} aims to accelerate research within the application of RL to chemical process control, offering a standardized platform for developing and testing RL-based control strategies in realistic chemical process environments.

Future work on {\sf PC-Gym} could expand its chemical process model library to include more diverse and complex chemical systems. This focus would call on the chemical process control community to contribute their models to improve the applicability of {\sf PC-Gym}. Additionally, only fixed policies trained with large evaluation budgets were evaluated in this work hence only demonstrating how an algorithm performs under an ideal case of evaluations. Further work could be done to implement an online learning feature in the {\sf PC-Gym} to allow the testing of algorithms where sample efficiency is paramount. Lastly, creating interfaces for real-time data streaming and hardware-in-the-loop testing would facilitate the transition from simulation to real-world implementation of RL-based control strategies.
\section{Acknowledgements}
Maximilian Bloor would like to acknowledge funding provided by the Engineering \& Physical Sciences Research Council, United Kingdom, through grant code  EP/W524323/1. Calvin Tsay acknowledges support from a BASF/Royal Academy of Engineering Senior Research Fellowship. José Torraca would like to acknowledge funding provided by the Coordenação de Aperfeiçoamento de Pessoal de Nível Superior - Brasil (CAPES) - Finance Code 001. The authors would like to acknowledge Panagiotis Petsagkourakis for productive discussions over the course of this work.  
\newpage

\printbibliography
\end{document}